\documentstyle[aas2pp4]{article}

\newcommand{\beq}{\begin{equation}}
\newcommand{\eeq}{\end{equation}}
\def\beqa{\begin{eqnarray}}
\def\eeqa{\end{eqnarray}}
\newcommand{\lsim}{\lesssim}
\newcommand{\gsim}{\gtrsim}

\begin{document}

\title{AN ISOCURVATURE CDM COSMOGONY. II.  OBSERVATIONAL TESTS}  

\author{P. J. E. Peebles}
\affil{Joseph Henry Laboratories, Princeton University,
Princeton, NJ 08544; pjep@pupgg.princeton.edu}
\authoremail{pjep@pupgg.princeton.edu}
 
\begin{abstract}

A companion paper presents a worked model for evolution through
inflation to initial conditions for an isocurvature model for
structure formation. It is shown here that the model is
consistent with the available observational
constraints that can be applied without the help of numerical
simulations. The model gives an acceptable fit to the second  
moments of the angular fluctuations in the thermal background
radiation and the second through fourth moments of the measured
large-scale fluctuations in galaxy counts, within the
possibly significant uncertainties in these measurements. The
cluster mass function requires a rather low but
observationally acceptable mass density, $0.1\lsim\Omega\lsim 0.2$ 
in a cosmologically flat universe. Galaxies would be assembled
earlier in this model than in the adiabatic version, an arguably
good thing. Aspects of the predicted non-Gaussian character
of the anisotropy of the thermal background radiation in this
model are discussed.

\end{abstract}
\keywords{cosmology: theory --- cosmology: large-scale structure
of universe --- galaxies: formation}

\section{Introduction}

An accompanying paper (Peebles~1998a; hereafter Paper~I) presents
a worked example of the evolution of a cosmological model through
inflation to initial conditions for an isocurvature (ICDM) model
for structure formation in a universe that now is dominated by
cold dark matter. Here I show that the model can be adjusted to
fit main observational constraints.

As in Paper~I, I attempt to keep the discussion simple and
definite by adopting a specific set of model parameters chosen
to give a reasonable approximation to the observations.
More detailed parameter studies that seek to minimize $\chi ^2$
measures of fit to the full suite of constraints would be
interesting but perhaps are not yet a pressing need because many
important observational constraints still are preliminary and may
harbor systematic errors. 

The adiabatic cold dark matter (ACDM) 
model for structure formation has been subject to 
searching tests from numerical simulations (eg. Governato {\it et
al.} 1998; Springel {\it et al.} 1998; and references therein). 
I hope the simpler observational tests presented here
show that the considerable effort needed for a meaningful
application of numerical simulations of the ICDM model would be
worthwhile.   

The model parameters are listed in \S 2. Second moments of
the angular distribution of the thermal background radiation (the
CBR) and the large-scale space distribution of galaxies
are presented in \S 3. In the ICDM model the primeval CDM mass
distribution is proportional to the square of a random Gaussian
process with zero mean. In \S 4 I discuss the nature of the
large-scale non-Gaussian fluctuations in the mass distribution
and compare them to third and fourth moments of galaxy
counts. The mass function 
of rich clusters of galaxies is discussed in \S 5. Because the 
distribution of mass fluctuations is broader than a Gaussian with
the same standard deviation, rare mass concentrations form earlier
than in an ACDM model. The ICDM model thus requires a lower mean
mass density for given normalization of the power spectrum, and
the cluster mass function changes significantly less rapidly with
redshift than in the ACDM model. In \S 6 I present the
scaling relation between the epochs of assembly of the dark
matter concentrations in galaxies and in rich clusters of
galaxies.  The relatively early assembly of protogalaxies 
in the ICDM model is arguably attractive. Finally, \S 7
presents some considerations of the higher moments of the angular
fluctuations of the CBR. As an example I compute the third moments
of the quadrupole and octupole components of the CBR anisotropy.
Concluding remarks are presented in \S 8.  

\section{Model Parameters}

The cosmological parameters are the same as in Paper~I, 
\beqa
	&&\Omega = 0.2, \qquad\Omega _{\rm B}\lsim 0.05,
	\qquad \lambda =0.8,\nonumber\\
	&&T_o=2.73\hbox{ K},\quad
	H_o=70\hbox{ km s}^{-1}\hbox{ Mpc}^{-1}.
\label{eq:cosmology}
\eeqa
The density parameter in baryons is $\Omega _{\rm B}$, and the
density parameter in CDM is $\Omega -\Omega _{\rm B}$. The model
is cosmologically flat; $\lambda$ is the fractional
contribution to the square of the expansion rate by a term in the
stress-energy tensor that is (or acts like) a cosmological
constant $\Lambda$.

The primeval entropy per baryon is a fixed universal value, to
agree with the standard model for the origin of the light
elements. There are no spacetime curvature fluctuations at high
redshift: the net mass density is homogeneous. Homogeneity is 
broken by the irregular primeval distribution of the CDM, which
is assumed to be a massive scalar field (or its decay remnants)
squeezed by inflation from its ground level to a classical field 
$\phi ({\bf x})$. The resulting mass distribution after inflation
ends and before the field starts oscillating (when the Hubble
length increases to the Compton wavelength of the CDM) is  
\beq
	\rho ({\bf x}) = m^2\phi ({\bf x})^2/2,
\eeq
where $\phi ({\rm x})$ is a random Gaussian process with zero
mean and power spectrum
\beq
	P_\phi\propto k^{m_\phi},\qquad m_\phi\sim -2.4, 
\label{eq:pphi}
\eeq
on scales of interest. The field autocorrelation function is
\beqa
	&& \xi _\phi (x_{12}) = 
	{\langle\phi ({\bf x_1})\phi ({\bf x_1})\rangle\over
	\langle\phi ^2\rangle}\nonumber\\
	&& =\int {d^3k\over (2\pi )^3}P_\phi 
	e^{i{\bf k}\cdot {\bf x}}\propto x_{12}^{-(3+m_\phi )},
		\label{eq:xiphi}
\eeqa
for the values of $m_\phi$ of interest here (eq.~[\ref{eq:pphi}]). As
discussed in \S 4, the mass autocorrelation function is 
\beq
	\xi = 2	{\langle\phi ({\bf x_1})\phi ({\bf x_1})\rangle ^2
	\over\langle\phi ^2\rangle ^2}\propto x_{12}^{-(6 + 2m_\phi )},
\label{eq:xirho}
\eeq
and the mass fluctuation power spectrum is
\beq
	P(k) = \int d^3x\xi (x)e^{i{\bf k}\cdot {\bf x}}
	\propto k^{m_\rho},
\eeq
where 
\beq
	m_\rho = 3 + 2m_\phi = -1.8
\label{eq:mphi}
\eeq
The numerical value fits the second moments of the CBR and
large-scale galaxy distributions (\S 3). The simplest inflation model
for isocurvature initial conditions gives $m_\phi =-3$; the model in
Paper~I is arranged to produce the wanted ``tilt'' in
equations~(\ref{eq:pphi}) and~(\ref{eq:mphi}). 

The power spectrum is normalized to 
\beq
	P(k) = 6300h^{-3}\hbox{ Mpc}^3 \hbox{ at } 
	k = 0.1h\hbox{ Mpc}^{-1},
\label{eq:Pnorm}
\eeq
where the adopted Hubble parameter is $h=0.7$
(eq.~[\ref{eq:cosmology}]). The rms fluctuation in the mass in a 
sphere of radius $8h^{-1}$~Mpc is $\sigma _8=0.9$. By this
traditional measure the model is biased, but as discussed next
the normalization fits one of the better measurements of the 
second moment of the large-scale galaxy distribution. 

Finally, the ionization history is computed in the standard way,
taking account of the slowing of recombination by the
Lyman-$\alpha$ recombination radiation, and under the assumption
that there is no significant source of ionizing radiation other
than the CBR and the recombination.

\section{Second Moments of the Distributions of Mass and
Radiation}

The spherical harmonic expansion of the angular distribution of
the CBR is
\beq
	T(\theta ,\phi) = \sum a_l^m Y_l^m(\theta ,\phi ),
\label{eq:alm}
\eeq
and the second moments may be written as
\beq
	T_l = \left[ {l(2l + 1)\over 4\pi } \right] ^{1/2}
	\langle |a_l^m|^2\rangle .
\label{eq:Tl}
\eeq
In the approximation of the sum over $l$ as an integral, 
$\sum l^{-1}\sim\int dl\, /l$, the variance of the
CBR temperature per 
logarithmic interval of $l$
is $(T_l)^2$.\footnote{There are good
historical reasons, dating from the introduction of the ACDM
model, for writing $2l(l+1)$ in place of $l(2l+1)$, but since I
am considering the ICDM case the use of the convention in
equation~(\ref{eq:Tl}) that takes account of the $2l+1$ values of
$m$ for given $l$ may be reasonable.} 

The ICDM model results in Figure~1 use the parameters in \S 2, and
show the effect of adjusting the baryon density parameter from
$\Omega _{\rm B}=0.05$ to 0.01. The measured $T_l$ are from the
compilation of Ratra (1998). The model is low at $l\sim 100$, and
perhaps also at $l=2$, depending on the correction for the
Galaxy. Gawiser and Silk (1998) present more detailed comparisons
to the measurements. I conclude that since these difficult
measurements may contain undetected systematic errors the model
fit is about as good as might be expected.

The power spectrum of the spatial distribution of mass in the
ICDM model is shown in Figure~2. 
The data are from the IRAS PSC-z (point source catalog) redshift
survey of Saunders {\it et al.} (1998). This is the real space
spectrum after correction for peculiar velocity distortion
represented by the density-bias parameter $\beta =0.6$. There are
good measurements of the 
spectrum of the galaxy distribution on smaller scales,
$k>0.1h$~Mpc$^{-1}$, but this approaches the nonlinear sector,
and it seems appropriate to postpone discussion of structure on
relatively small scales until we have more detailed explorations
of nonlinear evolution from the non-Gaussian initial conditions
of this model. Since the PSC-z catalog is deep, with good sky
coverage, it promises to be an excellent probe of the large-scale
galaxy distribution. Again, the model fit seems to be as good as
might be expected. 

\section{Higher Moments of the Galaxy Counts}

In the ICDM model the primeval CDM mass distribution is the
square of a Gaussian with zero mean value. I consider first some
statistical properties of the distribution and then compare
the moments to galaxy counts.

\subsection{Statistical Character of the Mass Fluctuations}

The mass density is 
$\rho ({\bf x})\propto\phi ({\bf x})^2$, where
$\langle\phi\rangle = 0$ and the autocorrelation
function of the Gaussian field is
\beq
	\langle\phi ({\bf x}_1)\phi ({\bf x}_2)\rangle
	\propto x_{12}^{-\epsilon},\quad \epsilon = 
	(m_\rho + 3)/2=0.6,
\label{eq:cphi}
\eeq
for the power spectrum index $m_\rho$ in
equation~(\ref{eq:mphi}). The fluctuation 
spectrum is suppressed on large scales where radiation pressure
cannot prevent the tendency of the isocurvature model to remain
homogeneous, and nonlinear evolution distorts the primeval
spectrum on small scales. Here I consider intermediate scales,
$k\sim 0.1h$~Mpc$^{-1}$, where the shape of the mass fluctuation
spectrum is close to the primeval form represented by
equation~(\ref{eq:cphi}).

The mass distribution smoothed through a window of volume $V$,
with the mean subtracted (and units chosen so the field mass is
unity), is 
\beq
	\delta\rho _s = \int {dV\over V}(\phi ({\bf x})^2
	- \langle\phi ^2\rangle ),
\eeq
and the $n^{\rm th}$ central moment is
\beq
	\langle (\delta\rho _s)^n\rangle = 
	\int {d^nV\over V^n}\langle\Pi(\phi ({\bf x_i})^2
	- \langle\phi ^2\rangle )\rangle.
\eeq
The integrand is the expectation value of a sum of products of
field values at the positions of the variables of integration 
${\bf x}_i$. Since $\phi ({\bf x})$ is a Gaussian
process the expectation value of each term is the sum of
products of two-point correlation functions 
$\langle\phi ({\bf x_i})\phi ({\bf x_j})\rangle$ for all ways of
pairing all positions $i,j$. All cases with $i=j$ are eliminated
by the term $\langle\phi ^2\rangle$, leaving the terms in
which every position appears in a two-point function with a different
position. That is, the integral is over a sum of terms each of
which is a product of $n$ factors $x_{ij}{}^\epsilon$ with
$i\not= j$. For $0<\epsilon <1.5$, as in  
equation~(\ref{eq:cphi}), the integrals converge at small
$x_{ij}$ and scale with the window size $r$ (the radius, if a
spherical window) as 
$\langle (\delta\rho _s)^n\rangle\propto r^{-n\epsilon}$.  
In particular, the variance scales as
$\langle (\delta\rho _s)^2\rangle\propto r^{-2\epsilon}$.  
The moments of the density contrast 
$\delta =\delta\rho _s/\langle\rho\rangle$ thus satisfy
\beq
	\langle(\delta /\sigma )^n\rangle = F_n,
	\qquad \sigma =\langle\delta {}^2\rangle ^{1/2},
\eeq
where $F_n$ is independent of the window size. That is, given the
window shape the probability distribution function of the ratio
$\delta/\sigma$ of the density contrast to its standard deviation
has a universal form, independent of the window size. 

I have not been able to find an analytic expression for the
distribution function of $\delta /\sigma$ for given $\epsilon$
and window shape, but can report numerical values for low moments.
The exercise of contracting the products of pairs of the 
$\phi ({\bf x}_i)$ in two-point functions yields the 
reduced spatial two-point correlation function 
(eqs.~[\ref{eq:xirho}], [\ref{eq:cphi}]),
\beq 
	\xi _2 = \langle\delta _1\delta _2\rangle 
		= 2(x_c/x_{12})^{2\epsilon},
\label{eq:xi2}
\eeq
where $x_c$ is a coherence length, the reduced
three-point function, 
\beq
   \xi _3 = \langle\delta _1\delta _2\delta _3\rangle = 
	8r_c^{3\epsilon}/(x_{12}x_{23}x_{31})^\epsilon,
\label{eq:xi3}
\eeq
and the reduced four-point function, 
\begin{eqnarray}
  \xi _4 & = &\langle\delta _1\delta _2\delta _3\delta _4\rangle 
 - \langle\delta _1\delta _2\rangle\langle\delta _3\delta _4\rangle
 - \langle\delta _1\delta _3\rangle\langle\delta _2\delta _4\rangle
		\nonumber\\
 && \qquad
 - \langle\delta _1\delta _4\rangle\langle\delta _2\delta _3\rangle
	\nonumber\\
 & = & 16r_c^{4\epsilon}[(x_{12}x_{23}x_{34}x_{41})^{-\epsilon}
	+(x_{12}x_{24}x_{43}x_{31})^{-\epsilon} \nonumber\\
 && \qquad
 + (x_{13}x_{32}x_{24}x_{41})^{-\epsilon}].\label{eq:xi4}
\end{eqnarray}
The skewness of the smoothed mass contrast satisfies 
\beq
	D_3 = \langle (\delta /\sigma )^3\rangle = 
	2^{3/2}{\langle (x_{12}x_{23}x_{31})^{-\epsilon}\rangle\over
	\langle x_{12}^{-2\epsilon}\rangle ^{3/2}}.
\label{eq:skewness}
\eeq
The angular brackets on the right-hand side mean the integral
over the window and divided by the window volume, a convenient
form for Monte Carlo integration. The excess kurtosis satisfies 
\beqa
	D_4 &=& \langle (\delta /\sigma )^4\rangle 
	-3\langle (\delta /\sigma )^2\rangle ^2 \nonumber\\
	&=& 12{\langle (x_{12}x_{23}x_{34}x_{41})^{-\epsilon}\rangle\over
	\langle x_{12}^{-2\epsilon}\rangle ^2}.
\label{eq:kurtosis}
\eeqa
Numerical values for  $\epsilon = 0.6$ and a square (top hat)
spherical window are 
\beqa
	D_3 &=& \langle (\delta /\sigma )^3\rangle = 2.46,\nonumber\\
	D_4 &=& \langle (\delta /\sigma )^4\rangle 
	-3\langle (\delta /\sigma )^2\rangle ^2 = 9.87.
\label{eq:eq20}
\eeqa

\subsection{Higher Moments of Deep Galaxy Counts}

Moments of galaxy counts in catalogs of angular positions
have been studied to test the theory of the onset of the
gravitational growth of nonlinear clustering out of an initially
Gaussian mass distribution. In the ICDM model there are initial
intrinsic higher order moments as well as what is generated by
the onset of  nonlinear clustering. This analysis takes account
of the intrinsic part only. The notation follows Peebles (1980),
with some adjustments to modern conventions. 

It will be assumed that a galaxy at distance $r$ is in the
angular catalog with probability proportional to a single 
selection function $\psi (r)$, and that the selection
probabilities along different lines of sight are statistically
independent. Then the reduced $n$-point angular correlation
function is  
\beqa
&& w_n(1,2,\ldots n) = \int\xi _n(1\ldots n)\,\Pi\psi _ir_i^2dr_i,
	\nonumber\\
	&& \int _0^\infty \psi (r) r^2dr = 1. 
\label{eq:wn}
\eeqa
The integral of the reduced spatial function, $\xi _n$, is over 
the volume element per steradian along each line of sight and
weighted by the selection function. The angular average of this
expression across a field of solid angle $\Omega$ is 
\beq
\bar w_n = \int \xi _n(1,\ldots n)\,\Pi\psi _i\, d^nV /\Omega ^n 
	= \langle\xi _n\rangle. 
\label{eq:wbar}
\eeq
As indicated in the last expression, this is an average
of the spatial correlation function over $n$ positions placed
uniformly at random in the angular field and with distributions in
radial positions given by 
\beq
	P(<r)=\int_0^r\psi (r) r^2dr,
\label{eq:Pofr}
\eeq
where $P$ is uniformly distributed from $P=0$ to $P=1$. Again,
this is a convenient form for Monte Carlo integration.  

With the correlation functions in equations~(\ref{eq:xi2})
to~(\ref{eq:xi4}) one sees that average of the three-point 
angular function (eq.~[\ref{eq:wbar}]) satisfies
\beq
	d_3 = {\bar w_3\over\bar w_2^{3/2}} =
	2^{3/2}{\langle (r_{12}r_{23}r_{31})^{-\epsilon}\rangle\over
	\langle r_{12}^{-2\epsilon}\rangle ^{3/2}},
\label{eq:d3}
\eeq
and the average of the fourth moment satisfies
\beq
	d_4 = {\bar w_4\over\bar w_2^2} = 
	 12{\langle (r_{12}r_{23}r_{34}r_{41})^{-\epsilon}\rangle\over
	\langle r_{12}^{-2\epsilon}\rangle ^2}.
\label{eq:d4}
\eeq
These expressions are averages through conical
windows, as in equations~(\ref{eq:skewness})
and~(\ref{eq:kurtosis}), but here the radial distribution
is given by equation~(\ref{eq:Pofr}).

The selection function may be modeled in terms of a Schechter
luminosity function,   
\beq
	\psi (r) \propto\int _{4\pi r^2f}^\infty
	\left( L\over L_\ast\right) ^\alpha {dL\over L_\ast}
	e^{-L/L_\ast },
\eeq
where the probability a galaxy is in the catalog is the
probability the energy flux density from the galaxy at the
observer exceeds the threshold $f$. This can be rewritten as 
\beq
	\psi\propto\int _{r/D_\ast}^\infty u^{1+2\alpha}e^{-u^2}du,
	\quad D_\ast = (L_\ast/4\pi f)^{1/2}.
\label{eq:psi}
\eeq
The characteristic depth, $D_\ast$, of the catalog does not
enter the scale-invariant ratios in equations~(\ref{eq:d3})
and~(\ref{eq:d4}). 

Table 1 shows numerical values of the ratios $d_3$ and $d_4$
(eqs~[\ref{eq:d3}] and~[\ref{eq:d4}]) computed using the
selection function in equation~(\ref{eq:psi}) with 
$\alpha = -1.0$. At $\alpha = -1.5$ the ratios $d_3$ and $d_4$ 
are about 10\%\ smaller. These ratios for narrow cones are
smaller than for a spherical window (eq.~[\ref{eq:eq20}]), but
the difference is not large. 

The relations between the means $\bar w_n$ of the correlation
functions and the moments of counts in cells is discussed in
Peebles (1980 \S 36) and in more generality by Gazta\~naga
(1994). In the former notation, and in terms of the central
moments  
\beq
	\mu _n=\langle (N-\langle N\rangle )^n\rangle ,
\eeq
for counts $N$ in cells, the third and fourth reduced moments
corrected for shot noise are
\beqa
	\langle N\rangle ^3\bar w_3 & = &
		\mu _3 - 3\mu _2 + 2\langle N\rangle ,\\
	\langle N\rangle ^4\bar w_4 & = &
	\mu _4 - 3\mu _2^2 -6\mu _3 + 11\mu _2 - 6 \langle N\rangle .
		\nonumber
\eeqa
Table~1 shows ratios of the $\bar w_i$ derived from galaxy counts 
in the Edinburgh-Durham Southern Galaxy Catalogue (Collins,
Nichol, \&\ Lumsden 1992) and the APM Catalogue (Maddox et al.
1990). The fields are square, with solid angle 
$\theta\times\theta$. In the analysis of the onset of the
gravitational growth of nonlinear clustering out of an initially
Gaussian mass distribution one is interested in the ratios
\beq
	s_n = \bar w_n/\bar w_2^{n - 1}.
\label{eq:sn}
\eeq
I convert the estimates of $s_n$ from the EDSGC catalog 
(Table~1 in Szapudi, Meiksin \&\ Nichol 1996) to the $d_n$ using
\beq
	\bar w_2=0.07\theta ^{-0.67},
\label{eq:w2bar}
\eeq
where $\theta$ is measured in degrees, from a power
law fit to the two-point
correlation function $w(\theta )$ at $0.1^\circ <\theta<2^\circ$
(Nichol \&\ Collins 1993). Gazta\~naga (1998) kindly provided the
$d_n$ derived from his analysis of moments of counts in the APM
catalog (Gazta\~naga 1994). 

Szapudi \&\ Gazta\~naga (1998) point out that the APM and EDSGC
catalogues are independently obtained from the same photographic
plates. They show that the moments of counts in the subsample of
APM in the smaller field of EDSGC are in satisfactory agreement
with the EDSGC moments. The substantial difference of the $d_3$
and $d_4$ at $\theta = 2^\circ$ thus is in the sky or plates or
more limited size of the EDSGC field. Szapudi \&\ Gazta\~naga 
note that since the APM field is substantially larger it likely
is the more reliable, but that can leave room for appreciable 
uncertainty in the moments from APM. 

A simpler measure may be relevant: the two-and three-point
spatial functions in equations~(\ref{eq:xi2}) and~(\ref{eq:xi3}) 
extrapolated to $x_c=x_{12}=x_{23}=x_{31}$ in linear theory
satisfy $\xi _2=2$, $\xi _3=8$, and $Q=0.7$, close to the
observed value in the hierarchical form for the galaxy
three-point function. Thus I conclude that the measurements are
not inconsistent with the model prediction under linear
perturbation theory. 

A more serious challenge comes from the correction for the
nonlinear growth of clustering. In the numerical
analysis of evolution from the non-Gaussian initial conditions of
a texture model Gazta\~naga \& M\"ah\"onen (1996) find the
skewness parameter grows from the initial value $D_3 = 0.7$ at  
$\sigma =\langle\delta ^2\rangle ^{1/2}= 0.1$ to $D_3 = 4$ 
at $\sigma = 1$. Comparable growth from the
initial value $D_3=2.5$ in the ICDM model (eq.~[\ref{eq:eq20}]) 
could make unacceptably large skewness, but that does depend on
the relative effect of the primeval skewness and excess kurtosis
on the growth of the variance and skewness. Potentially powerful
probes of possible departures from non-Gaussian initial conditons
from the properties of clusters of galaxies are discussed by
Chiu, Ostriker, \&\ Strauss (1998), and Robinson, Gawiser, \&\
Silk (1998). It
remains to be seen whether the initial conditions of the ICDM
model evolve to the observed clustering hierarchy of the galaxy
distribution at $\sigma\gsim 1$ (Scoccimarro et al. 1998; Fosalba
\&\ Gazta\~naga 1998).

\section{The Cluster Mass Function}

It is usually agreed that the Press-Schechter~(1974)
approximation offers a useful way to estimate the mass fraction in
rare concentrations such as rich clusters of galaxies when the
primeval density fluctuations are Gaussian. Here I apply a
variant of the Press-Schechter method based on a numerical
determination of the mass fraction in rare peaks in the ICDM
model.

The numerical determination starts with a realization of a random
Gaussian process with power law power spectrum in a $256^3$ cubic
grid of positions and wavenumbers. For the ICDM model the
realization is squared and then averaged on the lattice points through
a square (top hat) window of radius $r_s$ to get a smoothed
mass distribution. Peaks are defined as points on the lattice
where the density is larger than at the six nearest lattice
points. The mass fraction in peaks is the fraction of randomly
placed points that fall within distance $r_s$ of a peak with
smoothed density contrast $\delta _p$ larger than a chosen multiple of the
standard deviation, $\delta _p>\nu\sigma$. In the non-Gaussian
case the 
Gaussian process has power spectrum $P_\phi\propto k^{-2.4}$, so
the mass fluctuation spectrum is $P_\rho\propto k^{-1.8}$. In a
comparison Gaussian model the field is not squared, and here 
$P_\phi\propto k^{-1.8}$. The numerical results in Figure~3 are
based on the smoothing window radius $r_s = 4.1$, in units where
the box width is 256. Results at half the value of $r_s$ are
quite similar. An analytic fit for the ICDM model is
\beq
	f(>\delta /\sigma ) = 0.37 e^{-0.67\delta /\sigma }.
\eeq

Following the Press-Schechter method, the CDM mass distribution at
high redshift is smoothed through a spherical window that
contains the wanted mass, and the smoothed density contrast
is extrapolated to the epoch of interest in linear perturbation 
theory. The mass fraction in peaks in this distribution, with
density contrast 
\beq
	\delta _p = \nu\sigma _\rho >\delta _c = 1.68,
\label{eq:deltac}
\eeq
is the approximation to the mass fraction present in collapsed 
concentrations at the chosen epoch. Here $\delta _c$ is the
critical contrast for spherical collapse at $\Lambda =0$.

To get the present mass fraction in clusters I take the minimum
cluster mass to be
\beq
	m_{\rm cl} = 4\times 10^{14}h^{-1}M_\odot .\label{eq:mcl}
\eeq
The survey of Bahcall \&\ Cen (1993) indicates the present number
density of clusters at least this massive is
\beq
	n_{\rm cl} =(2\pm 1)\times 10^{-6}h^{-3}\hbox{ Mpc}^{-3}.
\eeq
 The cluster mass function varies roughly as
$n(>m)\propto m^{-2}$, so the mass fraction in clusters at
$m>m_{\rm cl}$ is
\beq
	f_{\rm cl} = 2m_{\rm cl}n_{\rm cl}/\langle\rho\rangle ,
\eeq
where $\langle\rho\rangle$ is the cosmic mean mass density. These
numbers give
\beq
	0.003<f_{\rm cl} \Omega <0.009.
\label{eq:fcl}
\eeq
The mass in equation~(\ref{eq:mcl}) is
contained in a sphere of comoving radius 
\beq
	r_{\rm cl} = 7.0h^{-1}\Omega ^{-1/3}\hbox{ Mpc},
\eeq
in the original near homogeneous mass distribution. The
mean square fractional fluctuation in galaxy counts in a
sphere of this radius in the present galaxy distribution (and
ignoring shot noise) is
reasonably well approximated by the power law model 
\beqa
\sigma _g{}^2 &=& 1.82[(4.5\pm 0.5)\hbox{ Mpc}/hr_{\rm cl}]^{1.77},
	\nonumber\\
\sigma _g &=& (0.91\pm 0.09)\Omega ^{0.30}.
\label{eq:sigmag}
\eeqa

To test the method I consider first the Gaussian case at
redshift $z=0$. For 
$\Omega =1$ the mass fraction in equation~(\ref{eq:fcl})
with the lower curve in Figure~3 translates to $3.1<\nu<3.5$.
The collapse condition in equation~(\ref{eq:deltac}) yields the
standard deviation in the mass contrast in the window that contains
$m_{\rm cl}$, $0.47<\sigma _\rho <0.54$. The ratio to the rms
fluctuation in galaxy counts from equation~(\ref{eq:sigmag}) is
\beq
	0.5 < \sigma _\rho/\sigma _g<0.7.
\label{eq:bhigh}
\eeq
For $\Omega = 0.2$ the same calculation for the Gaussian model
gives
\beq
	0.9 < \sigma _\rho/\sigma _g<1.4.
\label{eq:blow}
\eeq
Equations~(\ref{eq:bhigh}) and~(\ref{eq:blow}) are close to the
usual estimates of the bias needed to produce a reasonable
cluster mass function at the present epoch (Eke, Cole, \&\ Frenk
1996; Cen 1998). That is, this application of the Press-Schechter
procedure based on the mass fractions in peaks in a simulation
seems to be reasonably secure for the Gaussian case.  

For the non-Gaussian case the procedure using the upper curve in
Figure~3 with $\Omega =1$ gives
\beq
	0.2 < \sigma _\rho/\sigma _g<0.4.
\label{eq:42}
\eeq
For $\Omega = 0.2$, the value adopted in the ICDM model,
the method gives 
\beq
	3.1<\nu <4.7,
\label{eq:nunot}
\eeq
to fit the mass fraction in clusters, and 
\beq
	0.6 < \sigma _\rho/\sigma _g<1.0.
\label{eq:bbb}
\eeq
The more extended tail allows mass fluctuations at 
larger values of $\nu =\delta /\sigma$ than in the
Gaussian case. A result is that if $\Omega =1$ the suppression of
the fluctuations in mass relative to galaxies has to be even
stronger than in the Gaussian case. For $\Omega = 0.2$ 
equation~(\ref{eq:bbb}) indicates the ICDM model is just within
the bound from the cluster mass function and the assumption
galaxies trace mass, $\sigma _\rho=\sigma _g$.

The predicted evolution of the cluster mass function is much
slower in the non-Gaussian model than in the Gaussian case
because the broader tail of density fluctuations makes the mass
fraction a less sensitive function of $\delta /\sigma$. For
example, for $\Omega =0.2$ the growth factor for the density
contrast in linear perturbation theory from redshift $z=0.5$ to
the present is $D=1.24$. When $\nu$ in equation~(\ref{eq:nunot})
is multiplied by this factor it reduces the mass fraction 
$f_{\rm cl}$ (from the top curve in Fig.~3) by a factor of two, 
meaning the comoving cluster mass function at $z=0.5$ is half the
present value. The evolution is faster if $\Omega =1$: the
present 
central value $\nu = 6.2$ translates to $\nu = 9.3$ at $z=0.5$,
and the corresponding mass fraction is $f_{\rm cl}\sim 0.001$,
about 10 percent of the present value. This is not
inconsistent with the observations of clusters at $z\sim 0.5$,
but the strong bias is not attractive (eq.~[\ref{eq:42}]).  

\section{The Epoch of Galaxy Assembly}

In the CDM family of models, galaxies are assembled as mass
concentrations by a scaled version of the assembly of the mass
present now in rich clusters of galaxies. One sees from Figure~2
that in linear perturbation theory the mass fluctuation power
spectrum in the ICDM model is close to the primeval power law 
form at $k\gsim 0.1h$~Mpc$^{-1}$. Here the spectrum up to the
onset of development of nonlinear structure varies as
\beq
	P_\rho\propto k^{m_\rho}D(t)^2,
\label{eq:scaling}
\eeq
where $D(t)$ is the solution to the linear equation for the
evolution of the density contrast $\delta\rho /\rho$. The rms
contrast through a window of comoving radius $x$ thus scales as 
\beq
	\delta _s\propto x^{-(3+m_\rho )/2}D(t).
\eeq
Structure formation is triggered by
passage of upward fluctuations through $\delta _s\sim 1$, 
meaning the comoving length scale on which structure is
forming varies with time as 
\beq
	x_{\rm nl}\propto D^{2/(3+m_\rho )}.
\eeq
The corresponding physical length varies as
\beq
	r_{\rm nl}\propto (1+z)^{-1}D^{2/(3+m_\rho )},
\label{eq:rnl}
\eeq
the characteristic masses of newly forming objects is
\beq
	m_{\rm nl}\propto D^{6/(3+m_\rho )},
\label{eq:mnl}
\eeq
and the characteristic velocity dispersion within developing
structures is
\beq
	\sigma _{\rm nl}\propto (1+z)^{1/2}D^{2/(3+m_\rho )}.
\label{eq:vnl}
\eeq

These relations may be normalized to the great clusters of
galaxies, with
\beq
	r_{\rm A} = 1.5h^{-1}\hbox{ Mpc},\qquad
	\sigma _{\rm cl}=750\hbox{ km s}^{-1}.
\eeq
The line of sight velocity dispersion is an rms mean for 
$R\geq 1$ clusters. In the limiting isothermal sphere model these
numbers yield the mass in equation~(\ref{eq:mcl}). Clusters are
still relaxing at the Abell radius $r_{\rm A}$, and the merging
rate is not insignificant, but it is thought that the internal
velocities typically are close to what is needed for support
against gravity at $r\sim r_{\rm A}$. In the power law model in 
equation~(\ref{eq:scaling}) these quantities scaled back in time 
using equations~(\ref{eq:rnl}) to~(\ref{eq:vnl}) characterize
objects in a like state of early development in the past.  

With the parameters in equations~(\ref{eq:cosmology})
and~(\ref{eq:mphi}) the scaling relations applied at expansion
factor $1+z=7$ give 
\beqa
	&& r_{\rm g} = 20\hbox{ kpc},\
	\sigma _{\rm g} = 140\hbox{ km s}^{-1},\nonumber\\
	&& m_{\rm g} = 2\times10^{11}M_\odot .
\label{eq:young_galaxies}
\eeqa
This scaling calculation ignores the dissipative settling of the 
baryons to form the luminous central parts of the galaxies.
Consistent with this, in the standard model for an 
$L\sim L_\ast$ galaxy the mass within the radius $r\sim 20$~kpc
is dominated by dark matter.

In the ICDM model the large objects at $1+z=7$
with the parameters in equation~(\ref{eq:young_galaxies}) may be
compared to present-day clusters: there is significant    
substructure, the mass distribution in the outer parts is
disordered, and there is a significant rate of merging, but
the internal motions typically are close to what is needed for
virial support at radius $r\sim r_g$. The mass fraction in these
objects at $1+z=7$ is the same as the cluster mass
fraction from which they are scaled, 
$f_{\rm g}\sim (0.006\pm 0.003)/\Omega \sim 0.03\pm 0.015$
(eq.~[\ref{eq:fcl}]). The 
product of $f_{\rm g}$ with the mean mass density and divided by  
$m_{\rm g}$ gives a characteristic comoving number density,
\beq
	n_{\rm g}= (0.004\pm 0.002)h^3\hbox{ Mpc}^{-3}.
\label{eq:ngal}
\eeq
The characteristic size, internal velocity, mass, and number
density of these newly assembled systems are roughly typical of
present-day $L_\ast$ galaxies. An observer might be inclined to
call them young galaxies, assembled at $z\sim 6$. 

The scaling picture says the matter at $1+z=7$ that is not in
this generation of proto-galaxies would be in smaller clouds
between them, maybe positioned to become the Lyman-$\alpha$
forest observed at $z<5$.

At expansion factor $1+z=20$ the scaling relations give
\beqa
	&& r=1.3\hbox{ kpc},\ \sigma = 40\hbox{ km s}^{-1},
		\nonumber\\
	&& m=1\times10^{9}M_\odot ,
\eeqa
numbers characteristic of dwarf galaxies. I have to
postulate that some of these objects merge to contribute to the
mass  concentrations near the luminous parts of present-day giants.
In the model the generation of $L\sim L_\ast$
protogalaxies has been assembled at close to the present comoving
number density by the epoch $1+z=7$ (eq.~[\ref{eq:ngal}]), so I
must assume the rate of merging decreases, 
perhaps because the dissipative settling of the baryons has
progressed far enough to lower merging cross sections. 
Thereafter structure formation would build the present-day
galaxy  clustering hierarchy, while adding the extended massive
halos of galaxies at $r\sim 200$~kpc by accretion of 
relatively low mass star clusters and diffuse material.   
These numerous postulates could and should be checked by
numerical simulations.

Structure formation in the ACDM model is qualitatively the same
--- hierarchical growth by gravity --- but it happens at lower
redshift (Kauffmann 1996 and references therein). The remarkable
advances in the observations of 
high redshift young galaxies or their precursors may be bringing
us close to a test of these two picture. If galaxies were
assembled as mass concentrations at $1+z=7$ then the observations
at $1+z\simeq 4$ ought to reveal internal velocities
characteristic of present-day $L\sim L_\ast$ galaxies. This is
not inconsistent with the 
properties of the damped Lyman-$\alpha$ absorbers studied by
Wolfe \& Prochaska (1998), though Haehnelt, Steinmentz \&\ Rauch
(1998) show other
interpretations are possible. The expected optical appearance of
young galaxies at $1+z\sim 4$ depends on how feedback affects the
rate of conversion of gas to 
stars, a delicate issue that will require informed discussion.

\section{Higher Moments of the Thermal Background Radiation}

This discussion is limited to general remarks and an example, the
computation of the third moments of the quadrupole and octupole
parts of the anisotropy of the CBR predicted by the ICDM model. 

Because the statistical properties of the primeval mass
distribution in the model are more easily expressed in terms of
the position correlation functions than the power spectrum
and higher moments of the Fourier coefficients, I use a
Greens' function representation of the relation between the
primeval fluctuations in the CDM distribution and the observed
CBR anisotropy. With periodic boundary conditions the primeval
CDM mass distribution is
\beq
\delta ({\bf x}) = \sum \delta _{\bf k}e^{i{\bf k}\cdot {\bf x}},
\label{eq:deltap}
\eeq
and the spherical harmonic components of the observed CBR angular
distribution are (eq.~[\ref{eq:alm}]) 
\beq
	a_l^m = \sum _{\bf k}\delta _{\bf k}e_l(k)Y_l^m(\hat {\bf k}).
\label{eq:almp}
\eeq
Here $e_l(k)$ is the Legendre transform of the angular
distribution of the CBR produced by a single Fourier component
with wavenumber {\bf k} and normalized to amplitude 
$|\delta _{\bf k}|$ that is 
independent of $k$. The spherical harmonic is evaluated at the
direction  of {\bf k}. On expressing $\delta _{\bf k}$ in
equation~(\ref{eq:deltap}) as an integral over 
$\delta ({\bf x})$, and using
\beq
	e^{i{\bf k}\cdot {\bf x}} = 4\pi \sum _{l,m}i^lj_l(kx)
	Y_l^m(\hat {\bf k})Y_l^{-m}(\hat {\bf x}),
\eeq
one can reduce equation~(\ref{eq:almp}) to
\beq
	a_l^m = \int d^3x\delta ({\bf x})W_l(x)Y_l^m(\hat {\bf x}).
\label{eq:almpp}
\eeq
Here the spherical harmonic is a function of the direction of the
position {\bf x}, and the weight function is
\beq
	W_l(x) ={i^l\over 2\pi ^2}\int _0^\infty k^2dk\, e_l(k)j_l(kx).
\label{eq:wofl}
\eeq

The low order correlation functions of $\delta ({\bf x})$ in the
model are given by equations~(\ref{eq:xi2})
to~(\ref{eq:xi4}). Equation~(\ref{eq:almpp}) expresses
$a_l^m$  as an integral over $\delta ({\bf x})$, so it is
straightforward to write down expressions for the moments of the
$a_l^m$ as integrals over the spatial correlation functions. 

The third moments of the $a_l^m$ may be expressed in the form  
\beq
	S^{m_1,m_2,m_3}_{l_1,l_2,l_3} =
	{\langle a_{l_1}^{m_1}a_{l_2}^{m_2}a_{l_3}^{m_3}\rangle
	\over (\langle |a_{l_1}^{m_1}|^2\rangle
	\langle |a_{l_2}^{m_2}|^2\rangle
	\langle |a_{l_3}^{m_3}|^2\rangle )^{1/2}} =
	{N\over D}.
\label{eq:cbr_skewness}
\eeq
The numerator is
\beqa
	&&N=8\int d^3x_1d^3x_2d^3x_3\, (x_{12}x_{23}x_{31})^{-\epsilon}
		\nonumber\\
	&&\hspace{-9truemm} W_{l_1}(1)W_{l_2}(2)W_{l_3}(3)
	Y_{l_1}^{m_1}(1)Y_{l_2}^{m_2}(2)Y_{l_3}^{m_3}(3),
\label{eq:nform}
\eeqa
and the denominator is 
\beq
	D = (D_{l_1}D_{l_2}D_{l_3})^{1/2},
\eeq
where
\beqa
	&& D_l = 2\int d^3x_1d^3x_2\, x_{12}^{-2\epsilon}\nonumber\\
	&& \qquad
	W_{l}(1)W_{l}(2)Y_{l}^{m}(1)Y_{l}^{-m}(2).
\label{eq:dform}
\eeqa
Magueijo (1995) and  Ferreira, Magueijo, \& G\'orski (1998) point
out that the ratio of moments of the $a_l^m$ in
equation~(\ref{eq:cbr_skewness}) is a sensible way to normalize
this measure of the departure from Gaussian fluctuations in the
CBR. The ratio also very conveniently eliminates the
normalizations of the two- and three-point mass correlation 
functions, as in equation~(\ref{eq:d3}) for the skewness of the
distribution of galaxy counts. 

The integrals over angular positions are simplified by the
assumption that the density fluctuations are a stationary random
process, meaning the correlation functions of the 
$\delta ({\bf x})$ can only depend on relative 
positions. Thus in the expression for $D_l$ one can
average over all directions at fixed angular separation of the
lines of sight (1) and (2) using (Peebles 1973) 
\beq
	\langle Y_l^m(1)Y_l^{-m}(2)\rangle =
	P_l(\cos\theta _{12})/4\pi .
\label{eq:2average}
\eeq
(It would be even easier to go back to the expression for
$\langle |a_l^m|^2\rangle$ as an integral over the power
spectrum, but it seems best to compute numerator and denominator
in eq.~[\ref{eq:cbr_skewness}] in the same way.)

Luo (1994) shows one can similarly simplify the angle
integrals over the products of three spherical  
harmonics in equation~(\ref{eq:nform}). Because the $Y_l^m$ have
parity $(-1)^l$ it immediately follows that nonzero $N$ requires
\beq
	l_1+l_2+l_3 = \hbox{ even integer}.
\eeq
For even sums the symmetry of rotation
about the $z$-axis for given angular separations of the three
directions ${\bf x}_1$, ${\bf x}_2$, and ${\bf x}_3$
requires
\beq
	m_1 + m_2 + m_3 = 0.
\label{eq:mrule}
\eeq
As in quantum mechanics, the symmetry of rotation about any other
axis requires the triangle rule, 
\beq
	|l_1-l_2|\leq l_3\leq l_1+l_2.
\label{eq:lrule}
\eeq
To see this explicitly one can use the method in 
Peebles (1973, eq.~[B5]), which follows Edmonds (1957), to find
the mean of the product of three spherical harmonics averaged
over orientations for given angular separations of the three
lines of sight,  
\beq
 \langle Y_{l_1}^{m_1}(1)Y_{l_2}^{m_2}(2)Y_{l_3}^{m_3}(3)\rangle 
 = Z_{l_1,l_2,l_3}W_{m_1,m_2,m_3}^{l_1,l_2,l_3}
\label{eq:3average}
\eeq
where
\beqa
  && Z_{l_1,l_2,l_3} = \sum _m W_{0,m,-m}^{l_1,l_2,l_3}
		\nonumber\\
 && \qquad Y_{l_1}^0(0)Y_{l_2}^{m}(\theta _{12},0)
	Y_{l_3}^{-m}(\theta _{13},\phi _{23}).\label{eq:69}
\eeqa
The Wigner $3j$ symbols,
\beq 
	W_{m_1,m_2,m_3}^{l_1,l_2,l_3} = \left(\begin{array}{ccc}
  l_1 & l_2 & l_3 \\
  m_1 & m_2 & m_3 \\
	\end{array}\right) ,
\eeq
are proportional to Clebsch-Gordan coefficients, which vanish
unless the conditions in equations~(\ref{eq:mrule})
and~(\ref{eq:lrule}) are satisfied. 

In the isothermal CDM model the primeval fluctuations in the CDM
are balanced by opposing fluctuations in the radiation, so the
positive skewness in the CDM distribution is reflected in a
negative skewness in $\delta T/T$. The effect on the
distributions of the $a_l^m$ may be illustrated by the
expectation values of third moments for $l=2$
and $l=3$. Equation~(\ref{eq:3average}) (with tables of the
Wigner symbols, e.g. Edmonds 1957 Table 2) indicates the nonzero
skewness coefficients for $l=2$ in
equation~(\ref{eq:cbr_skewness}) satisfy 
\beqa
	 S_{2,2,2}^{0,2,-2}& = &(2/35)^{1/2}T_{2,2,2}, \nonumber\\ 
	 S_{2,2,2}^{0,1,-1} & = & (1/70)^{1/2}T_{2,2,2},\nonumber\\ 
	 S_{2,2,2}^{0,0,0} & = & -(2/35)^{1/2}T_{2,2,2}, \nonumber\\ 
	 S_{2,2,2}^{1,1,-2} & = & -(9/105)^{1/2}T_{2,2,2}.\label{eq:sratios}
\eeqa
These moments are unaffected by a permutation of the $m_i$
(eq.~[\ref{eq:cbr_skewness}]).
A sign change, $m_i\rightarrow -m_i$ for all $i$, produces the
complex conjugate of the product of observed $a_l^m$ and does not
affect the expectation value of this product. That is, one 
compares the expectation values to the real parts of the products
of $a_l^m$. A measurement of the $a_l^m$ from a full sky observation 
thus yields four measures of $T_{2,2,2}$, one of which is the
real part of $(a_2^1)^2a_2^{-2}$, which is the same as the real
part of  $(a_2^{-1})^2a_2^2$. The nonzero third moments involving
the $l=2$ and $l=3$ components are
\beqa
	 S_{3,3,2}^{0,0,0} & = & (2/105)^{1/2}T_{3,3,2},\nonumber\\
	 S_{3,3,2}^{1,-1,0} & = & -(3/140)^{1/2}T_{3,3,2},\nonumber\\
	 S_{3,3,2}^{3,-3,0} & = & (5/84)^{1/2}T_{3,3,2},\nonumber\\ 
	 S_{3,3,2}^{0,-1,1} & = & -(1/210)^{1/2}T_{3,3,2}, \nonumber\\
	 S_{3,3,2}^{1,-2,1} & = & (1/28)^{1/2}T_{3,3,2}, \nonumber\\
	 S_{3,3,2}^{2,-3,1} & = & -(5/84)^{1/2}T_{3,3,2}, \label{eq:mores}\\ 
	 S_{3,3,2}^{-1,-1,2} & = & (2/35)^{1/2}T_{3,3,2}, \nonumber\\
	 S_{3,3,2}^{0,-2,2} & = & -(1/21)^{1/2}T_{3,3,2}, \nonumber\\
	 S_{3,3,2}^{1,-3,2} & = & (1/42)^{1/2}T_{3,3,2}.\nonumber
\eeqa
These are nine measures of $T_{3,3,2}$.

A Monte Carlo numerical integration of
equations (\ref{eq:nform}) and~(\ref{eq:dform}) with
equations~(\ref{eq:2average}) and~(\ref{eq:3average})  
is straightforward (and the number of trial positions, 
$\sim 10^8$, needed for convergence no longer a problem). It is a
comforting check that the results for  
$T_{332}$ from $l_1=2$, $l_2=3$ and from $l_1=3$ and $l_2=2$
in equation~(\ref{eq:69}) agree. I find
\beq
	T_{2,2,2} = 1.02, \qquad T_{3,3,2} = -0.87.
\label{eq:ess}
\eeq

The analysis of third and fourth moments of other combinations
of low $l$ multipole components is straightforward in principle,
tedious to do by hand, perhaps easily done by 
computerized algebra, but beyond the scope of this paper. 
I expect that for significantly larger values of $l$
the best approach will be to analyze small sections of the sky in
rectangular coordinates that allow a much simpler representation
of the higher order moments.

In pioneering analyses Kogut et al (1996) found constraints on
some models for non-Gaussian fluctuations in the CBR angular
distribution, and Ferreira, Magueijo, \& G\'orski (1998)
found a set of measures of third moments of the $a_l^m$ in the
COBE DMR sky maps. I refrain from a comparison to 
equation~(\ref{eq:ess}) because the abbreviated
moment analysis presented here has little overlap with the 
measures obtained by Ferreira, Magueijo, \& G\'orski (1998). 

Because the initial mass distribution is homogeneous in the
isocurvature ICDM model there is no significant Sachs-Wolfe
effect at $l\lsim 30$; the CBR anisotropy simply reflects the
non-Gaussian fluctuations in initial composition. At smaller
scales radiation pressure is able to rearrange the net mass
distribution, producing significant contributions to the CBR
anisotropy from the Sachs-Wolfe effect and the motion of the 
baryons. This rearrangement shifts phases of Fourier components; 
thus a prominent hole in the initial distribution of baryons and
radiation becomes a ripple. I suspect this significantly
suppresses the ICDM prediction of non-Gaussian fluctuations in
the CBR at $l\gsim 100$. 

\section{Discussion}

\subsection{Is the ICDM Model Attractive from a Theoretical Point
of View?} 

Paper I compares the fields, parameters, and functional forms of
the potential energy in the ICDM model and other physical
theories for the seeds of structure from inflation. Here I
consider some broader issues. 

Our experience in particle physics might lead us to suspect that
the laws of physics relevant to the early universe will 
be found to be elegant and simple, albeit in some deeply subtle
way, and that once we understand the physics we will see that the 
universe is an expression of the physics. This world view informs 
many studies of inflation. The night thought of a physical
scientist might be that Newtonian mechanics is
expressed in fully developed turbulence, but it is not likely we
would know much 
about turbulence if we had not seen it. A knowledge of the
physics of the early universe might not be of much use if its
expression were complex.

Two themes could be accepted in either world view (as
well as by those, perhaps the majority, with more 
moderate opinions). First, the construction of a specific
internally consistent example of how evolution from very high
redshift could have led to the present state of the universe is
a valuable demonstration of consistency of the set of ideas
on which it is based. We have examples from the 
adiabatic CDM family of models. I have argued for yet another, 
an isocurvature CDM model. Second, the models and their
parameters will be reconsidered with each significant advance of
knowledge of the physics and astronomy, a process that will lead
us to abandon some models, adjust others, and maybe
introduce new ones. Perhaps this process will back us into
that narrow corner of model and parameter space that is a useful
approximation to what  really happened. We may have a modest
example in the fact that this latest version of the isocurvature
model (earlier steps of which may be traced back through Peebles
1997a) has the same dynamical actors as the ACDM model, though it
remains to be seen whether this is a lasting situation.

A mature physical theory must be falsifyable; there is
good reason for our conditioned dislike of theories that can be
adjusted to fit whatever is measured. On the other hand, if
the evolution of the early universe were moderately
complex we likely would need a flexible model to 
fit it. The isocurvature ICDM model in Paper I assumes power law
inflation because that makes it easy to select the fields and
their potential energy functions to produce a power law CDM
fluctuation spectrum $P_\rho\propto k^{m_\rho }$ over a wide range of
scales. But the evidence may lead us to another functional form.
In the power law ICDM models shown as the solid lines in
Figure~1 the CBR anisotropy $T_l$ at $l\sim 100$ may be too high
(Netterfield {\it et al.} 1997). That could be remedied by taking
the power law index $m_\rho$ to be closer to zero, but 
that would make $T_l$ unacceptably small at $l\lsim 10$. The
dotted line shows one way out: change the power spectrum to
\beq
	P_\rho\propto (k_x/k)^3 + (k_x/k)^{3/2},
\eeq
with 
\beq
	k_x = 0.01h\hbox{ Mpc}^{-1},
\eeq
and the other parameters the same as for the middle solid line. 
There has to be another bend to 
$P\propto k^{s}$,  $s<-3$, at $k\sim 1$~pc$^{-1}$ (Paper~I).
These bends are quite inelegant, unless Nature has chosen them.

The flexibility of the ICDM model is limited. For example, it is 
difficult to lower the spectrum at $l\sim 100$ without
significantly lowering the peak at $l\sim 300$. The advances in
observational constraints from work in progress will show whether
the ICDM model is a useful approximation. If the observations to
come in the next decade are fitted in all detail by one of the
simple structure formation models now under
discussion it will compel acceptance. If improving 
observations require increasingly baroque models it may mean we
have missed the correct elegant picture, or that the evolution of
the universe does not agree with our standards of elegance.

In my reading of the first of the world views mentioned above the
ICDM model is quite inelegant because it was constructed 
{\it ad hoc} to fit the observations and it is flexible enough to
be capable of adjustment to fit some substantial changes in the
observational situation. In the second world view, a model that
fits significant observational constraints within a sensible 
reading of the physics may not be all bad. 

\subsection{Is the Model Attractive from a Phenomenological Point
of View?}

The ICDM model has some possibly significant successes and
problems. Both will have to be
reconsidered with each advance of the observations and their
interpretation, of course. I hope is is not entirely self-serving
to note that a model that is close to reality may encounter
apparent problems as we sort out the ambiguities in the evidence.
Following are some considerations.  

\noindent i) The model parameters that
fit the CBR angular fluctuation spectrum $T_l$ in Figure~1
fit the second moment $P(k)$ of the
large-scale galaxy distribution in Figure~2, a significant
success. The low measured value of $T_l$ at $l\sim 100$ may
require adjustment of the model, or perhaps will prove be in some
part a systematic error in exceedingly difficult measurements.

\noindent ii) The skewness and excess kurtosis of galaxy counts
are not far from that of the model, a not insignificant result.
The major open issue is the correction for nonlinear evolution:
does the non-Gaussian primeval mass distribution of the model
evolve into the galaxy clustering hierarchy? An example of the
predicted non-Gaussian higher moments of the multipole expansion
components $a_l^m$ of the angular distribution of the CBR is
presented in \S 7. A comparison to the measurements remains to be
done.  

\noindent iii) The cluster mass function agrees with a
Press-Schechter approximation under the
assumption that galaxies trace mass at $\Omega = 0.2$. Chiu, 
Ostriker, \&\ Strauss (1998) point out that this
probes the nature of the primeval mass fluctuations, because
the cluster mass function depends on 
the tail of the distribution and the rms galaxy peculiar
velocity field on the standard deviation. The density parameter
in the ICDM model discussed here, $\Omega =0.2$, agrees with the
low peculiar velocities indicated by many analyses (e.g. Peebles 1986;
Bahcall, Lubin, \&\ Dorman 1995; Peebles 1997b; Willick \&\
Strauss 1998). The ICDM model thus seems to pass the Chiu et al.
test. If further work showed that the mass fraction $f_{\rm cl}$
in clusters is 
not near the upper end of the range in equation~(\ref{eq:fcl}),
and the Press-Schechter method is a
good approximation, it would require a lower value of $\Omega$,
increasing $T_l$ and tending to spoil the general consistency
with the measured $T_l$ and $P_\rho (k)$. 

\noindent iv) The model predicts relatively early galaxy
assembly; it is an open issue whether this is a success or
problem. The model may be considered a success from a theoretical
point of view because I arrived at it by a search for galaxy formation
at high redshift, when the mean mass density would have been
considerably closer to the relatively high density  
characteristic of the luminous parts of normal galaxies. The 
line of thought originated in Partridge \&\ Peebles (1967); a
recent version is in Peebles (1998b). The scaling
arguments in \S 6 suggest the ICDM model has some attractive
features as a model for galaxy formation. A more detailed
examination by numerical simulation remains to be done.  

My conclusion, from the second of the world views presented at
the beginning of this section, is that the ICDM model is
attractive because it fits a significant set of
observational constraints within what appears to be an acceptable
physical model.

\acknowledgements

I am grateful to Bill Ballinger and Will Saunders for 
the PSC-z data in Figure~2; to Neta Bahcall for guidance on
the cluster data; to Renyue Cen for his work on numerical
simulations that led to the analysis of
the cluster mass function in \S 5; to Avery Meiksin for help in
interpreting the EDSGC data in Table~1; to Enrique Gazta\~naga
for the APM data and stimulating discussions of the 
interpretation of the higher moments of the galaxy counts; and
to Bharat Ratra for the data compilation in Figure~1
and to him and Lyman Page for help in its interpretation. 
This work was supported in part by the National Science
Foundation.

\bigskip
\def\hrulefil{\leaders\hrule height0.6pt\hfill\quad}
\def\hrulefill{\leaders\hrule height0.6pt\hfill}
\centerline{TABLE 1}
\centerline{MOMENTS OF GALAXY COUNTS}
\hbox to \hsize{\hss\vbox{
\vskip 6pt 
\hrule height 0.6pt
\vskip 2pt
\hrule height 0.6pt
\halign{ \hfil #\hfil &\  \hfil #\hfil\  & \hfil #\hfil 
 &\  \hfil #\hfil\ 
&\  \hfil #\hfil\  & \hfil #\hfil  &\ \hfil # \hfil \cr
\noalign{\vskip 4pt}
\lower2ex\hbox{$\theta$} & 
\multispan{3} \hfil $d_3=\bar w_3/\bar w_2^{3/2}$\hfil & 
\multispan{3}\hfil $d_4=\bar w_4/\bar w_2^2$\hfil\cr
\noalign{\vskip -6pt}
 & \multispan{3} \hrulefil & \multispan{3} \hrulefill \cr
 & Model & EDSGC & APM & Model & EDSGC & APM \cr
\noalign{\vskip 4pt}
\noalign{\hrule height 0.6pt}
\noalign{\vskip 4pt}
0.5 & 1.07 & 1.5 & 1.4 & 2.60 & 4.4 & 4.7 \cr
1.0 & 1.25 & 1.2 & 1.0 & 3.40 & 3.1 & 2.2 \cr
2.0 & 1.49 & 1.4 & 0.7 & 4.35 & 4.0 & 1.2 \cr
\noalign{\vskip 4pt}
\noalign{\hrule height 0.6pt}
}}\hss}

\bigskip
\bigskip

\plotone{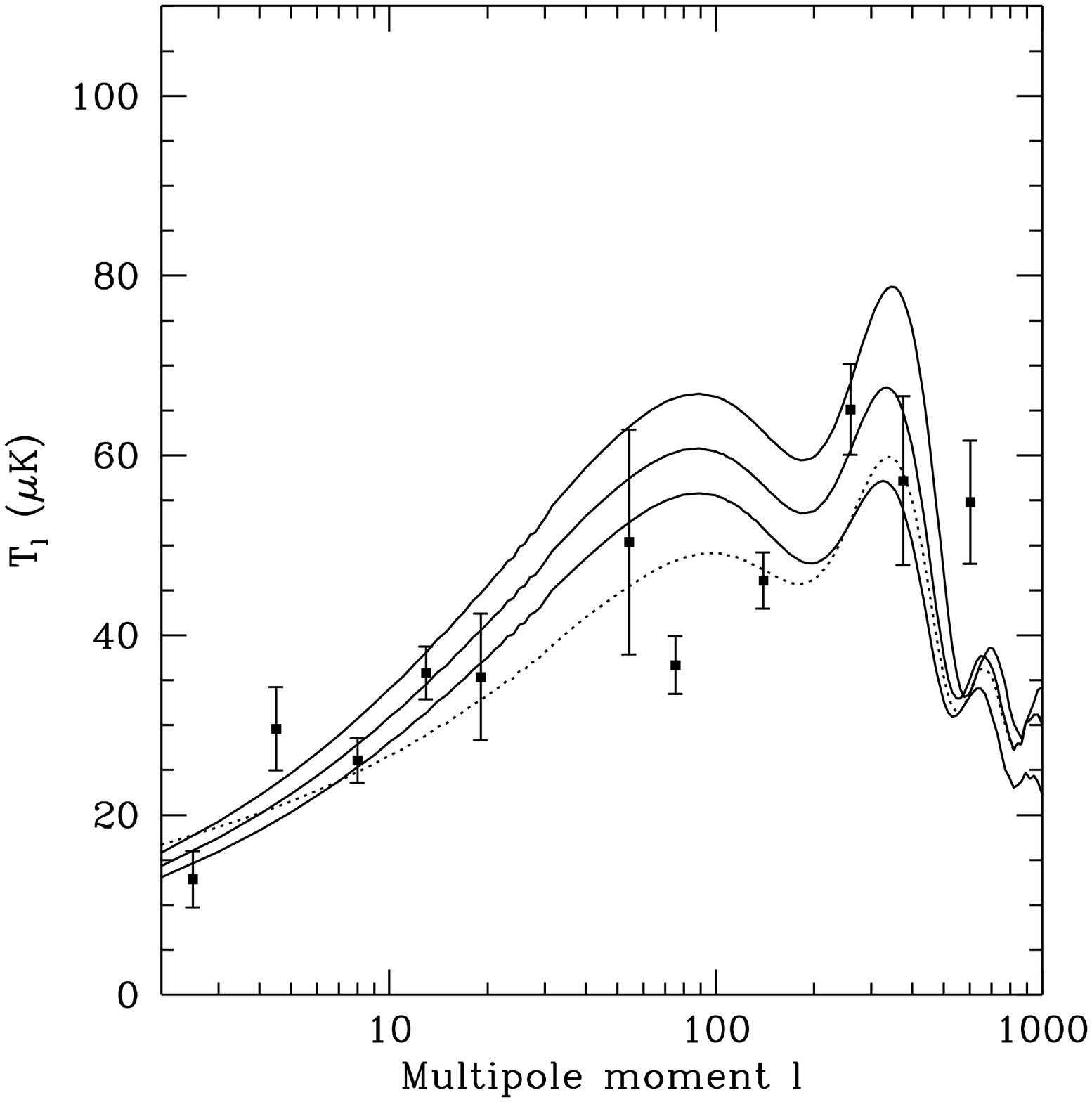}
\figcaption[figure1ii.ps]{Power spectrum of the CBR. The data
points are from the compilation by Ratra (1998). The ICDM model
assumes the parameters in equations~(\ref{eq:cosmology})
and~(\ref{eq:mphi}). The density parameter in baryons is 
$\Omega _{\rm B}=0.05$, 0.03, and 0.01 from the top to bottom
solid curve. The dotted curve is discussed in \S 8.}

\plotone{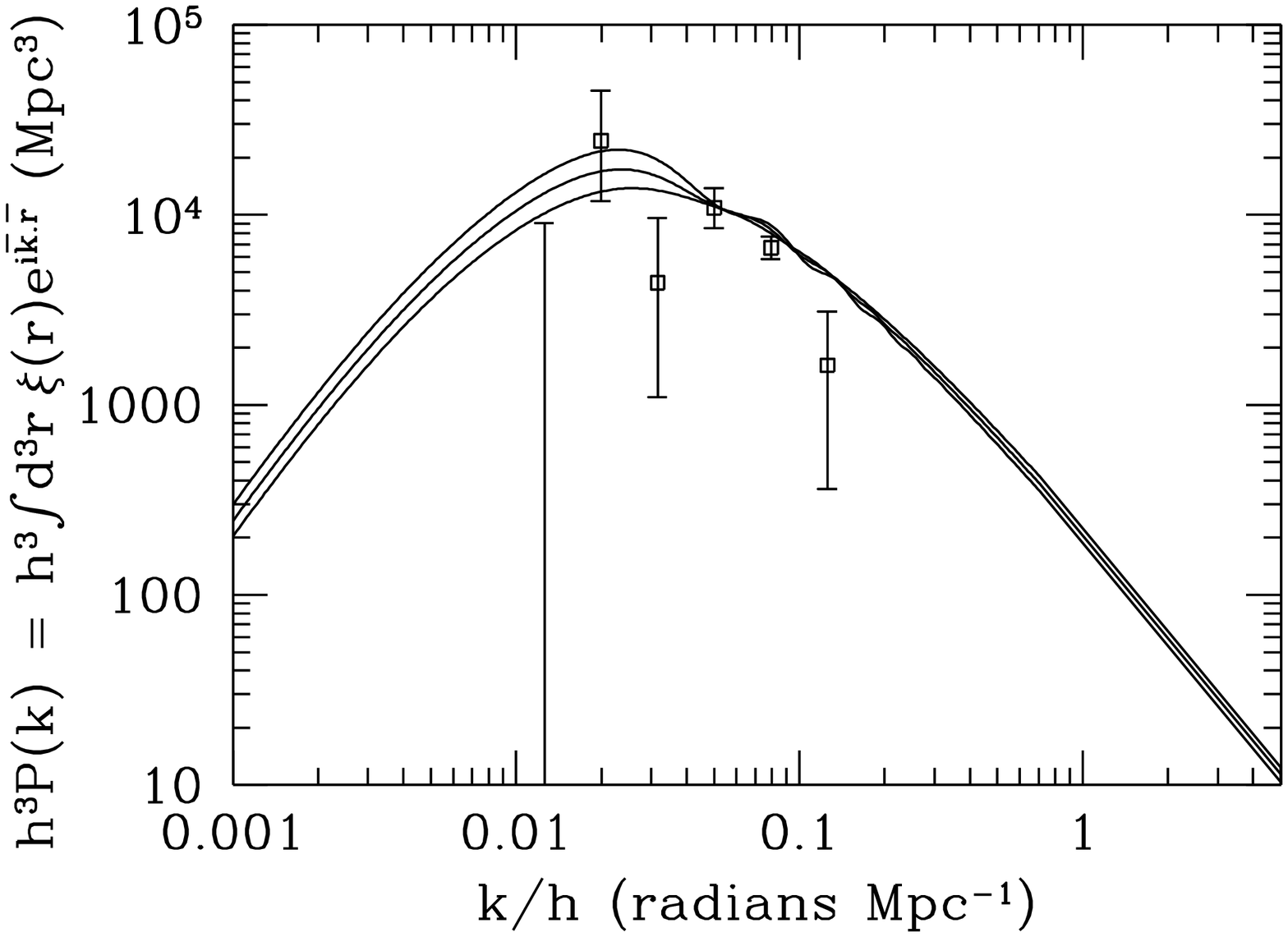}
\figcaption[figure2ii.ps]{Mass fluctuation spectra extrapolated to
the present in linear perturbation theory for the model in 
Fig.~1. The data are from the PSC-z
collaboration (Saunders {\it et al.} 1998). The density
parameter in baryons is  $\Omega _{\rm B}=0.05$, 0.03, and 0.01
from top to bottom at small wavenumber.} 

\plotone{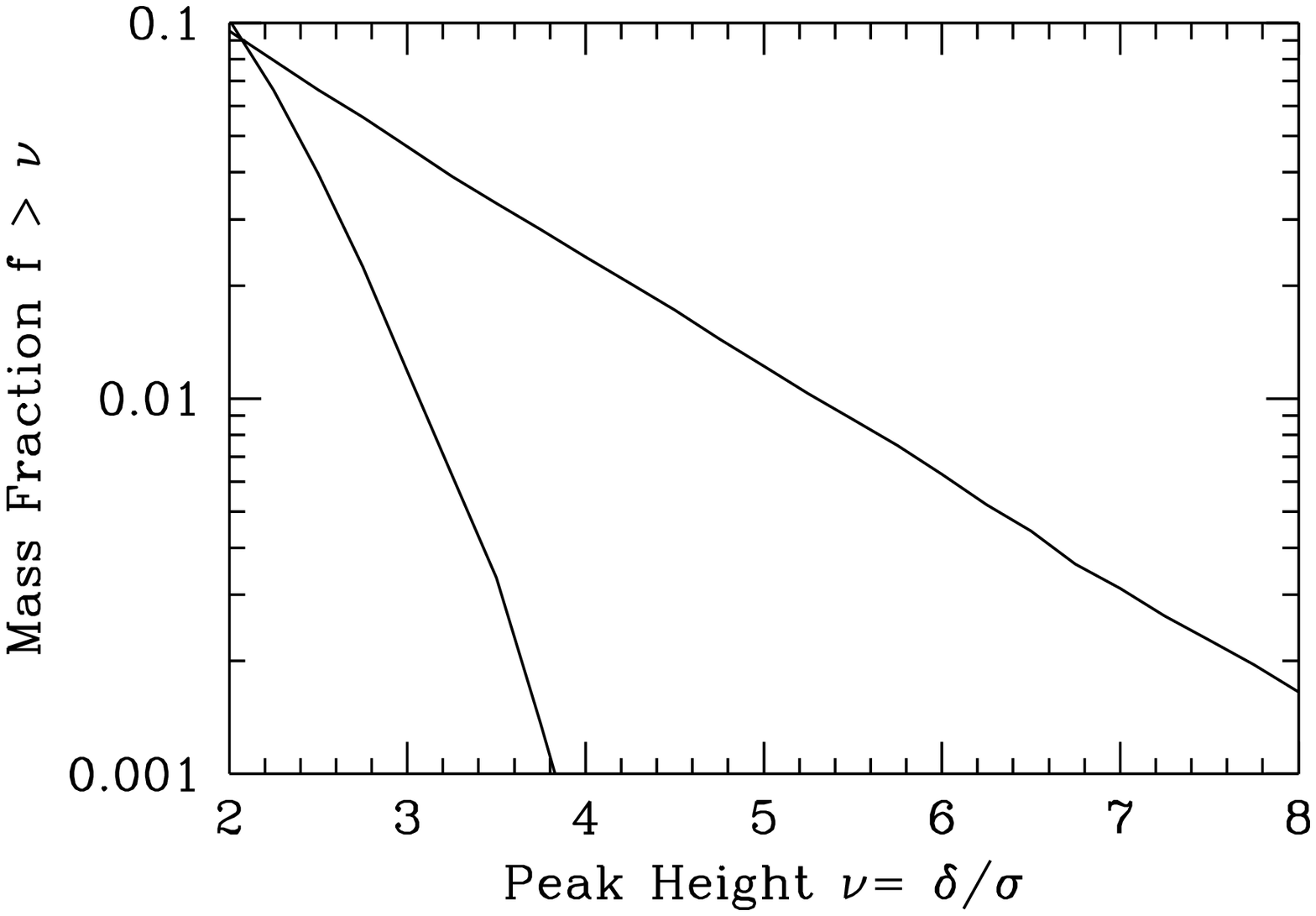}
\figcaption[figure3ii.ps]{Mass fractions in peaks at 
density contrast greater than $\nu$ standard deviations for a
Gaussian model (lower line) and the non-Gaussian ICDM model
(upper line). The models have mass power spectra $P\propto k^{-1.8}$. 
The mass distribution is smoothed through a spherical window, and
$f$ is the mass fraction within windows centered on peaks of the
mass distribution smoothed through the window.}


\begin{references}

\reference{net} Bahcall, N. A. \&\ Cen, R. 1993, ApJ, 407, L49

\reference{BLD} Bahcall, N. A., Lubin, L., \&\ Dorman, V. 1995,
	ApJ, 447, L81

\reference{Cen} Cen. R. 1998, preprint

\reference{Chiu} Chiu, W. A., Ostriker, J. P., \&\ Strauss. M. A. 
	1998, ApJ, 494, 479

\reference{EDSGC} Collins, C. A., Nichol, R. C., \&\ 
	Lumsden, S. L. 1992, MNRAS, 254, 295

\reference{Edmon} Edmonds, A. R. 1957, Angular Momentum in Quantum
Mechanics (Princeton: Princeton University Press)

\reference{Eke} Eke, V. R., Cole, S., \&\ Frenk, C. S. 1996,
	MNRAS, 282, 263

\reference{Fer} Ferreira, P. G., Magueijo, J. \& G\'orski, K. M.
	1998, preprint (astro-ph/9803256)

\reference{Fos} Fosalba, P. \&\ Gazta\~naga, E. 1998, 
	preprint (astro-ph/9802165)

\reference{Gaw} Gawiser, E. \&\ Silk, J. 1998, Science, in press

\reference{Gaz} Gazta\~naga, E. 1994, MNRAS, 268, 913

\reference{Gaz1} Gazta\~naga, E. 1998, private communication

\reference{Gaz2} Gazta\~naga, E. \&\ M\"ah\"onen, P. 1996, ApJ Lett
	462, L1

\reference{Sim1} Governato, F., Baugh, C. M., Frenk, C. S. 
Cole, S., Lacey, C. G., Quinn, T., \&\ Stadel, J. 1998, 
Nature, 392, 359

\reference{Hae} Haehnelt, M. G., Steinmentz, M. \&\ 
	Rauch, M. 1998, ApJ, 495, 647

\reference{Koff} Kauffmann, G. 1996, MNRAS, 281, 487

\reference{Kogut} Kogut, A., Banday, A. J., Bennett, C. L.,
	G\'orski, K. M., Hinshaw, G., Smoot, G. F.,
	\&\ Wright, E. L. 1996, ApJ, L29

\reference{Luo} Luo, X. 1994, ApJ, 427, L71

\reference{APM} Maddox, S. J., Sutherland, W. J., Efstathiou, G.,
	\& Loveday, L. 1990, MNRAS, 246, 433

\reference{Mag} Magueijo, J. C. R. 1995, Phys Lett, B342, 32

\reference{SK} Netterfield, C. B., Devlin, M. J., Jarosik, N.,
	Page, L., \&\ Wollack, E. J. 1997, ApJ, 474, 47

\reference{Nichol} Nichol, R. C. \&\ Collins, C. A. 1993, MNRAS,
	265, 867

\reference{PPP} Partridge, R. B. \&\ Peebles, P. J. E., 1967,
	ApJ, 147, 868

\reference{Peeb} Peebles, P. J. E. 1973, ApJ, 185, 413

\reference{Peebl} Peebles, P. J. E. 1980, The Large-Scale 
	Structure of the Universe (Princeton: Princeton 
	University Press)

\reference{Peebs} Peebles, P. J. E. 1986, Nature, 321, 27

\reference{Peebs} Peebles, P. J. E. 1997a, ApJ, 483, L1

\reference{Peebb} Peebles, P. J. E. 1997b, in Unsolved Problems
	in Astrophysics, eds. J. N. Bahcall and J. P Ostrker
	(Princeton: Princeton University Press), p. 1

\reference{Peeb1} Peebles, P. J. E. 1998a, preprint (Paper I)

\reference{Peeb2} Peebles, P. J. E. 1998b, Proc. NAS, 95, 67

\reference{PS} Press, W. H. and Schechter, P. 1974, ApJ, 187, 425 

\reference{Ratra} Ratra, B. 1998, private communication

\reference{Rob} Robinson, J., Gawiser, E., \&\ Silk, J. 1998,
	preprint

\reference{PSC-z} Saunders, W., {\it et al.} 1998, 
	in Extragalactic Astronomy in the Infrared, eds. G. A.
	Mamon, Trinh Xuan Thuan, \&\ J. Tran Thanh Van 
	(Gif-sur-Yvette: Editions Frontieres)

\reference{Sco} Scoccimarro, R., Colombi, S., Fry, J. N.,
	Frieman, J. A., Hivon, E., \&\ Melott, A., 1998, 
	ApJ, 496, 586

\reference{sim2} Springel, V., {\it et al.} 1998, submitted to
MNRAS (astro-ph/9710368)

\reference{Sz} Szapudi, I. \&\ Gazta\~naga, E. 1998, preprint
	(astro-ph/9712256)

\reference{Sza} Szapudi, I., Meiksin, A., \&\ Nichol, R. C. 1996,
	ApJ, 473, 15

\reference{Willick} Willick, J. A. \&\ Strauss, M. A. 1998, 
	preprint (astro-ph/9801307)

\reference{Wolfe} Wolfe, A. M. \&\ Prochaska, J. X. 1998, ApJ,
	494, L15

\end{references}
\end{document}